# Less-is-more in a 5-star rating system: an experimental study of human combined decisions in a multi-armed bandit problem

WATARU TOYOKAWA, KIM HYE-RIN AND TATSUYA KAMEDA, Hokkaido University

## 1. INTRODUCTION

A multi-armed bandit (MAB) problem grasps the exploration-exploitation dilemma, which is the fundamental trade-off in animal decision-making [Sutton & Barto, 1998; Gittins, 1979]. The decision maker must strike an optimal balance between taking an immediate reward by exploiting the option that has yielded the largest cumulative payoff and exploring other options to acquire more information about their expected payoffs. The majority of previous research has focused on comparing an individual decision performance against optimal policies in the context of individual reinforcement learning [Cohen, McClure & Yu, 2007].

Given the rapid proliferation of advanced information technologies, including the Internet, modern humans can easily access vast amount of socially transmitted information. Intuitively, this situation is isomorphic to some eusocial insects that are known to solve the exploration-exploitation dilemma collectively through information transfer (e.g., honeybees [Seeley et al., 1991]; and ants [Shaffer, Sasaki & Pratt, 2013]). Yet, in contrast from the eusocial insects, whose colonies are composed of kin, human collective performance may be affected by an inherent free-rider problem [Bolton & Harris, 1999; Kameda, Tsukasaki, Hastie & Berg, 2011]. Specifically, in groups involving non-kin members, it is expected that free-riders, who allow others to search for better alternatives and then exploit their findings through social learning ("information scroungers"), will frequently appear, and consequently undermine the advantage of collective intelligence [Rogers, 1998; Kameda & Nakanishi, 2003].

We focused on information transfer processes utilized by many of the "buzz-marketing" web sites, where consumers can learn about how other consumers have evaluated various products ("evaluation information", e.g., the 5-star rating system on Amazon.com), as well as how many others have purchased those products ("frequency information"). Does such information-sharing systems in fact improve performance in relation to the MAB problem? And, if so, how do the two types of social information, the frequency information and the evaluation information, affect our decisions?

We addressed this question through a laboratory experiment. A group of 5 participants were provided a series of thirty-armed bandit problems (Fig. 1). In each round, participants individually chose one option from 30 alternatives and received payoffs as personal rewards. In addition to the private payoff-feedback, information about other members' previous choice behaviors was publicly available during each round. We set up two conditions concerning the public information: (1) the 'frequency only' condition where individuals were informed of how many members had chosen each of the 30 alternatives in the preceding period; and (2) the 'frequency & evaluation' condition where, in addition to the social-frequency information, individuals could also learn other members' evaluations of their chosen-alternatives on a 5-point rating scale. As a baseline, we also had the individual condition where participants solved the task alone.

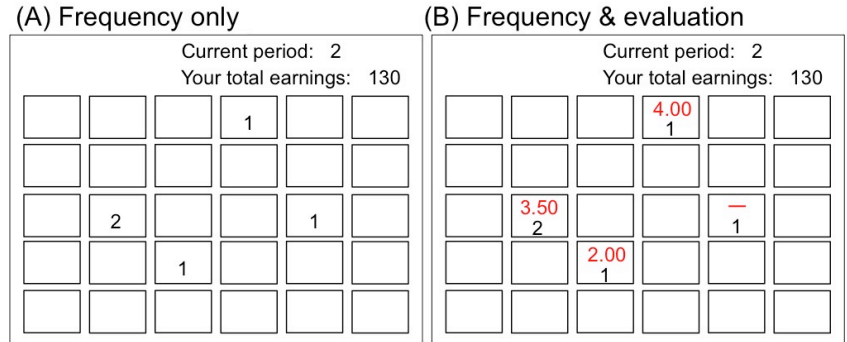

Fig. 1 Example displays of the task. After period 2, social information was shown at each choice stage. (A) An example of the frequency only condition. The numbers displayed within each box indicate the number of participants who chose this option for each box during the preceding round. (B) An example of the frequency & evaluation condition. In addition to the frequency information (black colored numbers), participants could see average ratings (red colored numbers) for each of the boxes that were evaluated during the previous round. The horizontal red bar indicated that no evaluation was contributed for that box during the preceding round. There was no social information feedback provided during the first round.

## 2. RESULTS SUMMARY

### 2.1  Did the information-sharing systems improve performance on the MAB problem?

- Yes. Fig. 2-a shows the total performances of social condition and individual condition.

### 2.2  Did the 5-point rating system (frequency & evaluation condition) improve performance compared with the frequency only condition?

- No. Fig. 2-b shows the time evolution of average performances of each condition. Statistical analysis indicated that the inclines of improvement on the average performance differed between two conditions, and in block 5, the average performance of frequency & evaluation condition was significantly lower than that of frequency only condition (frequency only condition: *mean performance* = 5.10; frequency & evaluation condition: *mean performance* = 4.49).

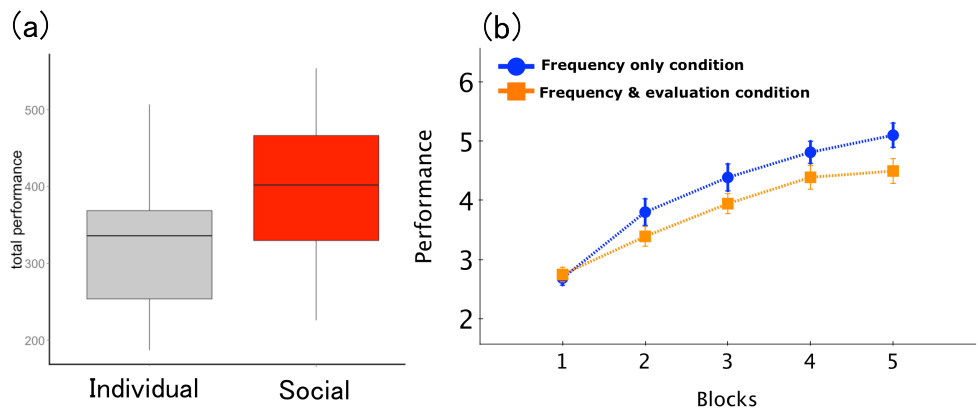

Fig. 2 (a): the total performance of the individual and social conditions. Participants earned 1 – 6 points each period depending on their choices. Therefore, the total performance could range 100 – 600 points. (b): the averaged performances of each condition of each block (20 rounds). The error bars show ± 1 SEM.

## 2.3 How did the 5-point rating system affect the performance?

- <u>"Less-is-more" effect emerged</u>. Fig. 3 shows a negative correlation between the total amounts of shared evaluations within a group and the total performance of each individual.

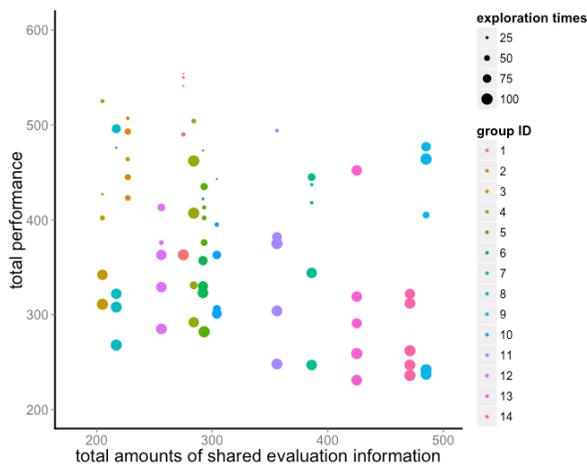

Fig.3 The horizontal axis shows the total amounts of shared evaluations within a group (minimum: 0, maximum: 500), and the vertical axis shows the total performance of each participant (minimum: 100, maximum: 600). Group identities are shown as different colors (participants belonging to the same group align on the same vertical line). The point size indicates exploration times of each participant (minimum: 1, maximum: 100).

## 3. DISCUSSION

Our results show that subjective evaluation information could undermine the benefit of collective intelligence through social learning in a multi-armed bandit problem. The average performance in the frequency only condition was higher than that in more informatized condition in which participants' subjective evaluations were also shared (Fig. 2-b). Moreover, in the frequency & evaluation condition, the amount of shared evaluation information was negatively correlated with individual total performance (Fig. 3).

     It has been argued that social influences sometimes undermine the effects of collective intelligence at simple estimation tasks [Lorenz et al. 2011]. In the estimation task of Lorenz et al. (2011), participants each had an opportunity to reconsider their response to factual questions after having received information about the responses of other participants. They concluded that social information had negative effects in terms of the accuracy of decision-making. Such simple estimation tasks have been broadly used in human collective intelligence studies [Galton, 1907; Krause et al. 2011]. However, such protocols have overlooked an important aspect of the nature of animal decision-making due to focusing only on the accuracy, and therefore failing to assess the cost-benefit trade-offs made through the whole decision processes. Since we would like to know the functional consequences of information transfer within a group, a framework that considers both the cost of information searching and the benefits of decision accuracy is required. Using the multi-armed bandit problem, we have provided the insight that the benefit of information transfer comes not only from improving the accuracy of decisions, but also in saving from the costs of exploration.

REFERENCES

Patrick Bolton and Christopher Harris. 1999. Strategic Experimentation. *Econometrica*. 67, 2 (1999), 349-374.

Jonathan D. Cohen, Samuel M. McClure and Angela J. Yu. 2007. Should I stay or should I go? How the human brain manages the trade-off between exploitation and exploration. *Phil. Trans. Roy. Soc. B*. 362, 933-942. DOI:http://dx.doi.org/10.1098/rstb.2007.2098

Francis Galton. 1907. Vox Populi. *Nature*. 75 (1907), 450-451.

John C. Gittins. 1979. Bandit processes and dynamic allocation indices. *J. Roy. Statist. Soc. B*. 41, 2 (1979), 148-177.

Tatsuya Kameda and Daisuke Nakanishi. 2003. Does social/cultural learning increase human adaptability? Rogers' question revisited. *Evolution and Human Behavior*. 24 (2003), 242-260.

Tatsuya Kameda, Takafumi Tsukasaki, Reid Hastie and Nathan Berg. 2011. Democracy under uncertainty: the wisdom of crowds and the free-rider problem in group decision making. *Psychological Review*. 118 (2011), 76-96.

Stefan Krause, Richard James, Jolyon J. Faria, Graeme D. Ruxton and Jens Krause. 2011. Swarm intelligence in humans: diversity can trump ability. *Animal Behavior*. 81 (2011), 941-948.

Jan Lorenz, Heiko Rauhutb, Frank Schweitzera, and Dirk Helbing. 2011. How social influence can undermine the wisdom of crowd effect. *Proc. Natl. Acad. Sci. USA*. 108 (2011), 9020-9025.

Thomas D. Seeley, Scott Camazine and James Sneyd. 1991. Collective decision-making in honey bees: how colonies choose among nectar sources. *Behavioral Ecology and Sciobiology*. 28 (1991), 277-290.

Zachary Shaffer, Takao Sasaki and Stephen C. Pratt. 2013. Linear recruitment leads to allocation and flexibility in collective foraging by ants. *Animal Behavior*. 37 (2013), 967-975.

Richard S. Sutton and Andrew G. Barto. 1998 *Reinforcement Learning: An Introduction (Adaptive Computation and Machine Learning)*. MIT Press, Cambridge, Massachusetts.

Alan R. Rogers. Does biology constrain culture? *American Anthropologist*. 90 (1988), 819-831.